\documentclass[12pt]{article}
\pdfoutput=1
\usepackage{graphicx}
\usepackage{hyperref}
\usepackage{epsfig}
\usepackage{amsmath}
\usepackage{amssymb}
\usepackage{color}
\usepackage[nosort]{cite}
\usepackage{hyperref}
\newlength{\abstractwidth}
\newcommand{\be}{\begin{equation}}
\newcommand{\ee}{\end{equation}}

\renewcommand{\title}[1]{\vbox{\center\bf{\Large{#1}}}\vspace{5mm}}
\renewcommand{\author}[1]{\vbox{\center#1}\vspace{5mm}}
\newcommand{\address}[1]{\vbox{\center\em#1}}
\newcommand{\email}[1]{\vbox{\center\tt#1}\vspace{5mm}}

\usepackage{mathrsfs} 

\usepackage{etoolbox} 
\usepackage{multirow}
\renewcommand\[{\begin{equation}}
\renewcommand\]{\end{equation}}

\newcommand{\ga}{\gamma}

\newcommand{\ba}{\begin{eqnarray}}
\newcommand{\ea}{\end{eqnarray}}

\hypersetup{pdfstartview=FitV,colorlinks=true,linkcolor=midblue,citecolor=midblue,filecolor=midblue,urlcolor=midblue}

\definecolor{midblue}{rgb}{0,0,0.5}

\begin{document}
	
	\begin{titlepage}
		\begin{center}
			\hfill \\
			\hfill \\
			\vskip 1cm
			
			\title{\Large \bf Light bending by a slowly rotating source\\ in quadratic theories of gravity}
			
			\author{\large Luca Buoninfante$^a$ and Breno L. Giacchini$^b$ }
			
			\address{{$^a$Department of Physics, Tokyo Institute of Technology, Tokyo 152-8551, Japan\\
			$^b$Department of Physics, Southern University of Science and Technology, Shenzhen 518055, China}}
			
			\email{\rm buoninfante.l.aa@m.titech.ac.jp,$\,$  breno@sustech.edu.cn}
			
			
			
		\end{center}

\begin{abstract}
In this paper we study the light bending caused by a slowly rotating source in the context of quadratic theories of gravity, in which the Einstein--Hilbert action is extended by additional terms quadratic in the curvature tensors. The deflection angle is computed employing the method based on the Gauss--Bonnet theorem and working in the approximation of a  weak lens; also, we assume that the source and observer are at an infinite distance. The formalism presented is very general and applies to any spacetime metric in the limit of weak gravitational field and slow rotation. We find the explicit formula for the deflection angle for several local and nonlocal theories, and also discuss some phenomenological implications. 
\end{abstract}

\end{titlepage}


\baselineskip=17.63pt


\section{Introduction}

Einstein's general relativity (GR) has gone through many challenges since its formulation, and its predictions have been verified to a very high degree of precision~\cite{-C.-M.}. At the same time, there are problems for which a satisfactory answer is still lacking: on galactic and cosmological scales, consistent descriptions for dark matter and dark energy have not been found yet. Moreover, in the short--distance (ultraviolet) regime, GR is plagued by cosmological and black hole singularities, whereas from the quantum point of view the theory is non--renormalizable~\cite{tHooftVeltman74}, losing predictability in the high--energy domain.

In the past decades such fundamental open questions have motivated many efforts towards a completion of GR. One of the most straightforward approaches is to generalize the Einstein--Hilbert action by including terms which contain fourth derivatives of the metric, such as $\mathcal{R}^2$, $\mathcal{R}_{\mu\nu}\mathcal{R}^{\mu\nu}$ and $\mathcal{R}_{\mu\nu\rho\sigma}\mathcal{R}^{\mu\nu\rho\sigma}$. The first interesting result in this context  traces back to~\cite{-K.-S.}, in which it was shown that the resultant theory is power--counting renormalizable. However, it still may not be regarded as a final theory because of the presence of a massive spin--$2$ ghost degree of freedom which causes Hamiltonian instabilities and breaks unitarity (when standard quantization prescriptions are implemented\footnote{See, \textit{e.g.}, Refs.~\cite{Anselmi,Anselmi:2017ygm,Donoghue:2019fcb,Salvio:2014soa,Salvio:2019wcp,Salvio:2019ewf,Chapiro:2019wua,Salles:2017xsr} for recent discussion on mechanisms through which fourth-order theories can be made unitary and stable.}).
A further important achievement of quadratic curvature gravity is given by the Starobinski model of inflation~\cite{starobinski}, based on the action extended by the term $\mathcal{R}^2$.

More recently, gravitational models with derivatives of order higher than four have also been intensively investigated. For example, GR--extended theories defined by actions with terms quadratic in the curvature tensors but with sixth and higher derivatives, like $\mathcal{R}\Box^n\mathcal{R}$ and $\mathcal{R}_{\mu\nu}\Box^n\mathcal{R}^{\mu\nu}$ ($n\geq 1$), can be super--renormalizable~\cite{Asorey:1996hz}. The unitarity of the $S$-matrix can also be restored in these models if the additional poles in the propagator appear as complex conjugate pairs~\cite{ModestoShapiro:2015ozb,Modesto:2015ozb,Anselmi:2017yux,Anselmi:2017lia}.

So far we have only mentioned examples of {\it local} quadratic theories of gravity, where the corresponding Lagrangian  depends polynomially on the derivative of the fields. Nonetheless, {\it nonlocal} modifications have also been proposed to deal with the aforementioned problems of renormalizability and unitarity~\cite{Krasnikov:1987yj,Kuzmin:1989sp,Tomboulis:1997gg,Biswas:2005qr,Modesto:2011kw,Biswas:2011ar}. In this case the Einstein--Hilbert action is enlarged by quadratic curvature terms such as $\mathcal{R}\mathcal{F}_1(\Box)\mathcal{R}$ and $\mathcal{R}_{\mu\nu}\mathcal{F}_2(\Box)\mathcal{R}^{\mu\nu}$, where $\mathcal{F}_1$ and $\mathcal{F}_2$ are non--polynomial functions. In particular, for specific choices of these functions one can have a ghost--free propagator and a renormalizable theory at the same time. Infrared modifications of GR can also be achieved, by means of non--analytic functions $\mathcal{F}_{1,2}$. For instance, the nonlocal operators $\mathcal{F}_i \propto \Box^{-1}$ and $\mathcal{F}_i \propto \Box^{-2}$ can produce an effect similar to the running of the Einstein--Hilbert term and the cosmological constant~\cite{GoSh}, respectively, and have fruitful applications in cosmology~\cite{Deser:2007jk,Maggiore2,NL-Z,Tan:2018bfp,Woodard:2018gfj,Maggiore1,Foffa:2013vma}.

All the models we mentioned above can be grouped under the label of \textit{quadratic theories of gravity}, as they are defined by an action of the form
\begin{equation}
\mathcal{S}=\frac{1}{2\kappa^2}\int {\rm d}^4x\sqrt{-g}\left\lbrace \mathcal{R}+\frac{1}{2} \left[ \mathcal{R}\mathcal{F}_1(\Box)\mathcal{R}+\mathcal{R}_{\mu\nu}\mathcal{F}_2(\Box)\mathcal{R}^{\mu\nu}+\mathcal{R}_{\mu\nu\rho\sigma}\mathcal{F}_3(\Box)\mathcal{R}^{\mu\nu\rho\sigma} \right] \right\rbrace\,,\label{quad-action}
\end{equation}
where $\kappa \equiv \sqrt{8\pi G}$ and the form factors $\mathcal{F}_i(\Box)$ are functions of the d'Alembertian $\Box$ which can be either local (polynomial) or nonlocal (non--polynomial).
Quadratic theories of gravity have been applied to several phenomenological contexts, and many observational constraints were derived~\cite{Deser:2007jk,Maggiore2,NL-Z,Tan:2018bfp,Woodard:2018gfj,Maggiore1,Foffa:2013vma,Conroy:2014eja,Deser:2013uya,Dodelson:2013sma,Capozziello:2011et,Blasone:2018nfy,Blasone:2018ftv,Buoninfante:2018bkc,Buoninfante:2019der,Buoninfante:2019uwo,Lambiase:2016bjy,Lambiase:2015yia,Capozziello:2014mea,Capozziello06,Clifton,Capozziello11,Berry-Gair,Stabile&Stabile,LWLM,Zhao:2017jmv,Finch:2016bum,Capozziello:2014rva,Capozziello:2020dvd,Napolitano:2012fp,Giacchini16,Accioly:2016etf,Accioly:2015fka,Accioly98,Accioly:2016qeb,Giacchini:2018twk}.

In this paper we aim to extend the results of the works~\cite{Accioly98,Accioly:2016qeb,Giacchini:2018twk}, where the light bending caused by a weak, static gravitational field was studied for particular classes of quadratic theories of gravity, namely, those with fourth and sixth derivatives, and those with $\mathcal{F}_2=\mathcal{F}_3 \equiv 0$. The generalization we present here is twofold. In what concerns the source of the gravitational field, we assume that it has a non--zero angular momentum and analyse the effect of its slow rotation on the deflection of light.
Moreover, we consider a wider class of gravitational theories, presenting explicit calculations for some important particular cases of nonlocal models and the general polynomial--derivative model. The formalism used to compute the deflection angle is based on the Gauss--Bonnet theorem~\cite{Gibbons:2008rj} and it follows the developments of~\cite{Ono:2017pie}, applicable to axisymmetric spacetimes. Our presentation is very general, in the sense that it applies to any model of the form~\eqref{quad-action} in the linear regime around Minkowski.

The paper is organized as follows. In Section~\ref{quad-theories} we review the framework of quadratic theories of gravity on the linear regime. We find the minimal set of differential equations for the unknown independent components of a general axisymmetric metric describing the surrounding spacetime of a slowly rotating source. In Section~\ref{se-light-bend}, we present the formalism to study the light bending in a rotating spacetime metric in the linear regime, or in other words, in the limit of weak gravitational field and slow rotation.
In Section~\ref{applic}, we apply such a general formalism to specific theories
and obtain an explicit expression for the corresponding deflection angle.
Section~\ref{conclus} is devoted to the discussion about the phenomenological implications of the results and conclusions.
Throughout the paper we adopt the mostly positive convention for the metric signature, $\mathrm{diag}\left(-,+,+,+\right),$ and the natural units system, $c=1=\hbar.$

\section{Quadratic theories of gravity in the weak--field limit}\label{quad-theories}

As our goal is to describe the deflection undergone by a light ray in a weak gravitational field, in this Section we review the main classical aspects of quadratic theories of gravity with particular focus on the linearised metric solutions for a slowly rotating source.

\subsection{Linearised field equations}

In the weak--field approximation we expand Eq.~\eqref{quad-action} around the Minkowski background,
\begin{equation}
g_{\mu\nu}=\eta_{\mu\nu}+\kappa h_{\mu\nu}\,,\label{lin-metric}
\end{equation}
where $h_{\mu\nu}$ is the metric perturbation, and keep only terms up to order $\mathcal{O}(h^2)$ in the action. Therefore, in the linear regime  one can simplify the action~\eqref{quad-action} by neglecting the term $\mathcal{R}_{\mu\nu\rho\sigma}\mathcal{F}_3(\Box)\mathcal{R}^{\mu\nu\rho\sigma}$. In fact, for analytic quadratic gravity, because of the identity
\begin{equation*}
\mathcal{R}_{\mu\nu\rho\sigma}\Box^n\mathcal{R}^{\mu\nu\rho\sigma}=4\mathcal{R}_{\mu\nu}\Box^n\mathcal{R}^{\mu\nu}-\mathcal{R}\Box^n\mathcal{R}+\mathcal{O}(\mathcal{R}^3)+{\rm div}\,,
\end{equation*}
where {\rm div} means total derivative terms, up to order $\mathcal{O}(h^2)$ the Riemann--squared contributions can be replaced by combinations of Ricci scalar and Ricci tensor squared (as $\mathcal{O}(\mathcal{R}^3)$ only contributes at order $\mathcal{O}(h^3)$). In what concerns non--analytic models, defined, \textit{e.g.}, by form factors proportional to $\Box^{-1}$ or $\Box^{-2}$, the direct expansion of the action using~\eqref{lin-metric} shows that also in this case the terms of order $\mathcal{O}(h^2)$ originated from the Riemann--squared term can be rewritten as a redefinition of the form factors $\mathcal{F}_1$ and $\mathcal{F}_2$ (see,\textit{ e.g.},~\cite{Conroy:2014eja} for an explicit calculation).
Hence, by working in the linear regime hereafter we set $\mathcal{F}_3(\Box)=0$ without any loss of generality.

The bilinear form of the action~\eqref{quad-action} reads~\cite{Biswas:2011ar}:
\begin{equation}
\begin{array}{rl}
\mathcal{S}^{(2)}=&\displaystyle \frac{1}{4}\int {\rm d}^4x\left\lbrace \frac{1}{2}h_{\mu\nu}f(\Box)\Box h^{\mu\nu}-h_{\mu}^{\sigma}f(\Box)\partial_{\sigma}\partial_{\nu}h^{\mu\nu}+hg(\Box)\partial_{\mu}\partial_{\nu}h^{\mu\nu}\right.\\[3mm]
& \,\,\,\,\,\,\,\,\,\,\,\,\,\,\,\,\,\,\,\,\,\,\,\,\,\,\,\,\,\,\displaystyle \left. -\frac{1}{2}hg(\Box)\Box h +\frac{1}{2}h^{\lambda\sigma}\frac{f(\Box)-g(\Box)}{\Box}\partial_{\lambda}\partial_{\sigma}\partial_{\mu}\partial_{\nu}h^{\mu\nu}\right\rbrace \,,
\label{lin-quad-action}
\end{array}
\end{equation}
where $h\equiv\eta_{\mu\nu}h^{\mu\nu}$  defines the trace and 
\begin{equation}
\begin{array}{rl}
f(\Box)\equiv & \displaystyle  1+\frac{1}{2}\mathcal{F}_2(\Box)\Box,\\
g(\Box)\equiv & \displaystyle 1-2\mathcal{F}_1(\Box)\Box-\frac{1}{2}\mathcal{F}_2(\Box)\Box\,.
\label{f-g}
\end{array}
\end{equation}
By variating the action in Eq.~\eqref{lin-quad-action} with respect to the field $h_{\mu\nu}$ it follows the corresponding linearised field equations,
\begin{equation}
\begin{array}{ll}
\displaystyle f(\Box)\left(\Box h_{\mu\nu}-\partial_{\sigma}\partial_{\nu}h_{\mu}^{\sigma}-\partial_{\sigma}\partial_{\mu}h_{\nu}^{\sigma}\right) \displaystyle +g(\Box)\left(\eta_{\mu\nu}\partial_{\rho}\partial_{\sigma}h^{\rho\sigma}+\partial_{\mu}\partial_{\nu}h-\eta_{\mu\nu}\Box h\right)&\\[3mm]
\,\,\,\,\,\,\,\,\,\,\,\,\,\,\,\,\,\,\,\,\,\,\,\,\,\,\,\,\,\,\,\,\,\,\,\,\,\,\,\,\,\,\,\,\,\,\,\,\,\,\,\,\,\,\,\,\,\,\,\,\,\,\,\,\,\,\,\,\,\,+\displaystyle \frac{f(\Box)-g(\Box)}{\Box}\partial_{\mu}\partial_{\nu}\partial_{\rho}\partial_{\sigma}h^{\rho\sigma}=-2\kappa T_{\mu\nu}\,,&
\label{lin-field-eq}
\end{array}
\end{equation}
where 
\begin{equation}
T_{\mu\nu}=-\frac{2}{\sqrt{-g}}\frac{\delta S_m}{\delta g^{\mu\nu}}\simeq \frac{2}{\kappa}\frac{\delta S_m}{\delta h^{\mu\nu}}
\end{equation}
is the matter stress--energy tensor sourcing the gravitational field, with $S_m$ being the action describing the matter sector, and it satisfies the conservation law $\partial^{\mu}T_{\mu\nu}=0$ consistently with the Bianchi identity.

We are interested in finding the linearised metric generated by a slowly rotating source,
\begin{equation}
{\rm d}s^2=-(1+2\Phi){\rm d}t^2+2\vec{h}\cdot {\rm d}\vec{r} \, {\rm d}t +(1-2\Psi)({\rm d}r^2+r^2{\rm d}\Omega^2)\,,\label{isotr-metric}
\end{equation}
where $r=\sqrt{x^2+y^2+z^2}$ is the isotropic radial coordinate and ${\rm d}r^2+r^2{\rm d}\Omega^2={\rm d}x^2+{\rm d}y^2+{\rm d}z^2,$ while $\kappa h_{00}=-2\Phi,$ $\kappa h_{ij}=-2\Psi\delta_{ij}$ and $\kappa h_{0i}=h_i$ are the metric potentials generated by a non--diagonal stress--energy tensor $T_{\mu\nu}.$ In particular, we assume that the source is pressureless, $T\equiv \eta^{\mu\nu}T_{\mu\nu}\simeq -T_{00},$ therefore its non--vanishing components are $T_{00}=\rho(r)$ and $T_{0i},$ \textit{i.e.}, the source is modelled as a rotating dust of density $\rho(r)$.

By making the assumption of stationary source and using the $00$--component and the trace of the linearised field equations~\eqref{lin-field-eq}, one can show that the metric potentials in Eq.~\eqref{isotr-metric} are the solutions of the following differential equations\footnote{We point out that in order to obtain the equation for $h_{i}$ in such a form it is necessary to impose the de Donder gauge condition $\partial_\mu h^\mu_\nu=0$ or a suitable higher--order generalization~\cite{Accioly:2016qeb,Teyssandier89,Holscher:2018jhm}, compatible with the metric~\eqref{isotr-metric}.}
\begin{equation}
\begin{array}{rl}
\displaystyle \frac{f(\nabla^2)[f(\nabla^2)-3g(\nabla^2)]}{f(\nabla^2)-2g(\nabla^2)}\,\nabla^2\Phi(r)=&\kappa^2 \,T_{00}(r)\,,\\[3mm]
\displaystyle \frac{f(\nabla^2)[f(\nabla^2)-3g(\nabla^2)]}{g(\nabla^2)}\,\nabla^2\Psi(r)=&- \kappa^2\,T_{00}(r)\,,\\[3mm]
\displaystyle f(\nabla^2)\,\nabla^2 h_{i}(r)=&- 2\kappa^2\,T_{0i}(r)\,,
\end{array}\label{field-eq-pot}
\end{equation}
where $f$ and $g$ are now functions of the Laplacian as $\Box\simeq \nabla^2$.
For $f=g=1,$ which also means $\mathcal{F}_1=\mathcal{F}_2=0,$ the differential equations in Eq.~\eqref{field-eq-pot} reduce to standard Poisson equations, matching the linearised limit of GR.

\subsection{General linearised metric solution for a slowly rotating source}\label{SecGenSol}

The differential equations for the metric potentials can be formally solved by finding the Green's functions and using the method of Fourier transform. Indeed, from~\eqref{field-eq-pot} we obtain:
\begin{equation} 
\Phi(r)= \displaystyle -2G\int {\rm d}^3r'\mathcal{G}_\Phi(\vec{r}-\vec{r}')T_{00}(\vec{r}')\,,
\label{fourier-pot-phi}
\end{equation}
\begin{equation}
\!\!\!\!\!\Psi(r)= \displaystyle 2G\int {\rm d}^3r'\mathcal{G}_\Psi(\vec{r}-\vec{r}')T_{00}(\vec{r}')\,,
\label{fourier-pot-psi}
\end{equation}
\begin{equation}
\!\!\!\!\!\!\!\!\!h_{i}(r)= \displaystyle 4 G\int {\rm d}^3r'\mathcal{G}_\zeta(\vec{r}-\vec{r}')T_{0i}(\vec{r}')\,,
\label{fourier-pot-non-diag}
\end{equation}
where the integration region is defined by the volume of the gravitational source, while  $\mathcal{G}_\ell(\vec{r}-\vec{r}')$ (with $\ell=\Phi,\Psi,\zeta$) are the Green's functions which eventually will only depend on the modulus $|\vec{r}-\vec{r}'|$ and are defined via
\begin{equation}
\!\!\!\!\!\!\!\!\!\!\!\!\!\!\!\!\!\!\!\!\!\!\!\!\!\!\!\!\!\!\!\!\!\!\!\!\!\!\!\!\!\!\!\!\!\!\frac{f(\nabla^2)[f(\nabla^2)-3g(\nabla^2)]}{f(\nabla^2)-2g(\nabla^2)}\nabla^2\mathcal{G}_\Phi(\vec{r}-\vec{r}')=-4\pi \delta^{(3)}(\vec{r}-\vec{r}')\,,
\label{green-func-def-1}
\end{equation}
\begin{equation}
\!\!\!\!\!\!\!\!\!\!\!\!\!\!\!\!\!\!\!\!\!\!\!\!\!\!\!\!\!\!\!\!\!\!\!\!\!\!\!\!\!\!\!\!\!\!\frac{f(\nabla^2)[f(\nabla^2)-3g(\nabla^2)]}{g(\nabla^2)}\nabla^2\mathcal{G}_\Psi(\vec{r}-\vec{r}')=-4\pi \delta^{(3)}(\vec{r}-\vec{r}')\,,
\label{green-func-def-2}
\end{equation}
\begin{equation}
f(\nabla^2)\nabla^2\mathcal{G}_\zeta(\vec{r}-\vec{r}')=-4\pi \delta^{(3)}(\vec{r}-\vec{r}')\,.
\label{green-func-def-3}
\end{equation}
Since we assume that the deflection of light is produced in a weak field regime, far outside the gravitational source, we can perform a multipole expansion,
\begin{equation}
\begin{array}{rl}
\displaystyle \mathcal{G}_\ell(|\vec{r}-\vec{r}'|)=&\displaystyle  \mathcal{G}_\ell(r)+\partial_j'\left.\mathcal{G}_\ell(|\vec{r}-\vec{r}'|)\right|_{r'=0}x'^j+\cdots\\[3mm]
=& \displaystyle \mathcal{G}_\ell(r)-\frac{1}{r}\frac{\partial \mathcal{G}_\ell(r)}{\partial r}x'_jx^j+\cdots\,,
\end{array}
\label{multip-green}
\end{equation}
where the ellipses stand for higher order multipole contributions. In the case of theories defined by analytic form factors $\mathcal{F}(\Box)$, \textit{e.g.}, in local and nonlocal higher--derivative gravities, $\mathcal{G}_\ell(r)$ can be computed by using the method of Fourier transform.
For theories defined by non--analytic form factors $\mathcal{F}_i\propto\Box^{-1}$ or $\mathcal{F}_i\propto\Box^{-2}$ it is necessary to fix suitable boundary conditions in order to define the corresponding Green's functions (see,\textit{ e.g.},~\cite{Maggiore1,Foffa:2013vma,Conroy:2014eja} for further discussion).

Let us first consider the diagonal components $\Phi$ and $\Psi$ and subsequently the cross--term $h_{i}.$ By using the expansion in Eq.~\eqref{multip-green} up to the dipole term, the first non--vanishing contributions for the diagonal components are
\begin{equation}
\displaystyle  \Phi(r)= \displaystyle   - 2G\,\mathcal{G}_\Phi(r)\int {\rm d}^3r'T_{00}(\vec{r}')=-2GM\, \mathcal{G}_\Phi(r)\,, \label{phi-sol} 
\end{equation}
\begin{equation}
\!\!\!\!\!\!\!\!\!\!\displaystyle  \Psi(r)= \displaystyle    2G\,\mathcal{G}_\Psi(r)\int {\rm d}^3r'T_{00}(\vec{r}')=2GM\, \mathcal{G}_\Psi(r)\,, \label{psi-sol} 
\end{equation}
where we have defined the mass of the system as
\begin{equation}
M=\int {\rm d}^3r'T_{00}(\vec{r}')\,. \label{mass} 
\end{equation}

As for the off-diagonal components, we can proceed as done for the diagonal part but
the first non--vanishing contribution will come from the dipole term. Indeed, the solution in Eq.~\eqref{fourier-pot-non-diag} can be rewritten as
\begin{equation}
\begin{array}{rl}
\displaystyle  h_{i}(r)= & \displaystyle  \displaystyle 4 G\,\mathcal{G}_\zeta(r)\int {\rm d}^3r'T_{0i}(\vec{r}')-\frac{4G}{r}\frac{\partial \mathcal{G}_\zeta(r)}{\partial r}x_j\int {\rm d}^3r' x'^jT_{0i}(\vec{r}')\\[3mm]
=& \displaystyle  -\frac{2G}{r}\frac{\partial \mathcal{G}_\zeta(r)}{\partial r}(\vec{r}\wedge \vec{J})_i\,,
\end{array}
\label{mult-non-diag}
\end{equation}
where we used the property\footnote{This relation is a consequence of the continuity equation $\partial_{\mu}T^{\mu\nu}=0.$ Indeed, one can easily show that $$\int {\rm d}^3r'T^{\mu i}(\vec{r}')=\int {\rm d}^3r'T^{\mu k}(\vec{r}')\delta_k^i=\int {\rm d}^3r'T^{\mu k}(\vec{r}')\frac{\partial x'^i}{\partial x'^k}=-\int {\rm d}^3r'\left(\partial_k' T^{\mu k}(\vec{r}')\right)x'^i=0.$$} $\int {\rm d}^3r' T_{0i}(\vec{r}')=0$ and introduced the angular momentum of the source through
\begin{equation}
\int {\rm d}^3r' T_{0i}(\vec{r}')x'^j=\frac{1}{2}\varepsilon_{ijk}J^k\,.
\label{angul-moment}
\end{equation}
Hence, we have found a formal expression for the cross--term in the metric in Eq.~\eqref{isotr-metric}. By choosing the direction of angular momentum along the $z$--axis, $\vec{J}=J\hat{z},$ and making the coordinate transformations $x=r\,{\rm sin}\theta\,{\rm cos}\varphi$ and $y=r\,{\rm sin}\theta\,{\rm sin}\varphi,$ we can write 
\begin{equation}
\begin{array}{rl}
\displaystyle 2\vec{h}\cdot d\vec{r}dt=&\displaystyle-\frac{4G}{r}\,\frac{\partial \mathcal{G}_\zeta(r)}{\partial r}(\vec{r}\wedge \vec{J})_i\,dx^idt\\[3mm]
\equiv & \displaystyle 2 \zeta(r)\,{\rm sin}^2\theta\,d\varphi\,dt\,,
\end{array}
\label{cross--term}
\end{equation}
where we have defined 
\begin{equation}
\zeta(r)\equiv 2GJ\,r\,\frac{\partial \mathcal{G}_\zeta(r)}{\partial r}\,.
\label{chi}
\end{equation}
Therefore, the spacetime metric in Eq.~\eqref{isotr-metric} can be recast in the form
\begin{equation}
{\rm d}s^2=-(1+2\Phi){\rm d}t^2+2\zeta(r)\,{\rm sin}^2\theta\,{\rm d}\varphi\,{\rm d}t +(1-2\Psi)({\rm d}r^2+r^2{\rm d}\Omega^2)\,,\label{polar-metric}
\end{equation}
where the metric potentials $\Phi,$ $\Psi$ and $\zeta$ can be found by using the expressions in Eqs.~\eqref{phi-sol}, \eqref{psi-sol} and \eqref{chi}. Note that in Einstein's GR we have $f=g=1$ which implies $\mathcal{G}_\Phi(r)=-\mathcal{G}_\Psi(r)=1/(2r)$ and $ \mathcal{G}_\zeta(r)=1/r,$ thus recovering the weak--field limit of the metric potentials for the Kerr metric~\cite{Kerr:1963ud}, \textit{i.e.}, the metric in Eq.~\eqref{polar-metric} would reduce to the Lense--Thirring form~\cite{Lense}.

\subsection{Field redefinition: spin--$2$ and spin--$0$ potentials}

As we shall see explicitly in the next Section (see also, \textit{e.g.},~\cite{Accioly98,Accioly:2016qeb,Giacchini:2018twk}), in computing the gravitational deflection undergone by a light ray it turns out that the spin--$2$ part of the propagator has a more prominent role, in the linear regime. It is convenient, thus, to work with the auxiliary potentials $\chi_{0,2}$ introduced in~\cite{Giacchini:2018gxp}, defined as
\begin{equation}
\chi_0\equiv \Phi-2\Psi \qquad \textrm{and} \qquad   \chi_2\equiv \Phi+\Psi  \,, 
\end{equation}
from which one can re--obtain the original ones as
\ba
\Phi= \frac{1}{3} (2 \chi_2 + \chi_0) \qquad \textrm{and} \qquad \Psi = \frac{1}{3} ( \chi_2 - \chi_0).
\ea
As a consequence, the field equations~\eqref{field-eq-pot} reduce to 
\ba
f_s(\nabla^2) \nabla^2 \chi_s & = & \kappa^2 T_{00}\,,\label{chi_s diff}
\\[3mm]
f_2(\nabla^2) \nabla^2 h_i & = & -2 \kappa^2 T_{0i}\,,\label{h_i diff}
\ea
where $s=0,2$ and we have defined\footnote{The propagator around Minkowski background is given by~\cite{Biswas:2011ar,Accioly:2002tz}
\begin{equation} \label{Propagator}
\Pi_{\mu\nu\rho\sigma}(k)=\frac{\mathcal{P}^2_{\mu\nu\rho\sigma}}{f_2(-k^2)k^2}+\frac{\mathcal{P}^{0-s}_{\mu\nu\rho\sigma}}{f_0(-k^2)k^2}\,,
\end{equation}
where $\mathcal{P}^2_{\mu\nu\rho\sigma}$ and $\mathcal{P}^{0-s}_{\mu\nu\rho\sigma}$ are operators that project along the spin--$2$ and spin--$0$ components~\cite{Barnes-Rivers,VanNieuwenhuizen:1973fi}, and terms which are gauge--dependent have been omitted. From the last expression it is clear that the field potentials $\chi_0$ and $\chi_2$ are associated to the spin--$0$ and spin--$2$ components, respectively.
} $f_2(z)\equiv f(z)$ and $f_0(z) \equiv f(z) - 3g(z)$. It follows that the three metric potentials are determined by two Green's functions $\mathcal{G}_{0,2}$, which are solution of 
\ba
f_s(\nabla^2) \nabla^2 \mathcal{G}_s(\vec{r}-\vec{r}')=-4\pi \delta^{(3)}(\vec{r}-\vec{r}')\,,\qquad s=0,2\,.
\ea
In this more economic notation, the Green's functions introduced in Eq.~\eqref{fourier-pot-phi}--\eqref{fourier-pot-non-diag} are related to $\mathcal{G}_{0,2}$ through
\ba
\mathcal{G}_0=\mathcal{G}_\Phi + 2\mathcal{G}_\Psi , \qquad  \mathcal{G}_2 = \mathcal{G}_\Phi - \mathcal{G}_\Psi = \mathcal{G}_\zeta .
\ea
Hereafter, for simplicity, we shall work only with the notation $\mathcal{G}_s$, where $s=0,2$.

In this spirit, the set of Eqs.~\eqref{fourier-pot-phi},~\eqref{fourier-pot-psi} and~\eqref{fourier-pot-non-diag} are equivalent to
\ba
\chi_s(r) & = & -2 G \int {\rm d}^3r'\mathcal{G}_s(\vec{r}-\vec{r}')T_{00}(\vec{r}')\,,\qquad s=0,2\label{int-chi_s}
\\[3mm]
h_i(r) & = & 4 G \int {\rm d}^3r'\mathcal{G}_2(\vec{r}-\vec{r}')T_{0i}(\vec{r}')\,.\label{int-h_i}
\ea
Similarly to the expressions for the potentials $\Phi$ and $\Psi$ in Eqs.~\eqref{phi-sol} and~\eqref{psi-sol}, up to the dipole term one has
\begin{equation}
\chi_0=-2GM\, \mathcal{G}_0(r)\,, \qquad  \chi_2=-2GM\, \mathcal{G}_2(r)\,.
\label{new potentials}
\end{equation}
Furthermore, the identification $\mathcal{G}_\zeta(r) = \mathcal{G}_2(r)$ allows us to write (always in the dipole approximation)
\begin{equation}
\zeta(r)\equiv - a r \, \frac{\partial \chi_2}{\partial r}\,,
\label{new potentials zeta}
\end{equation}
where we introduced the rotational parameter $a \equiv J/M$. 

Finally, in terms of the spin--0 and spin--2 potentials the metric~\eqref{polar-metric} reads
\begin{equation}
{\rm d}s^2=-\left[ 1+\frac{2}{3} (2 \chi_2 + \chi_0)\right] {\rm d}t^2-2 a r \chi_2^\prime\,{\rm sin}^2\theta\,{\rm d}\varphi\,{\rm d}t +\left[ 1-\frac{2}{3} ( \chi_2 - \chi_0)\right] ({\rm d}r^2+r^2{\rm d}\Omega^2)\,.\label{polar-metric2}
\end{equation}
Therefore, once the functions $f_{0,2}$ (or, in an equivalent manner, the form factors $\mathcal{F}_{1,2}$) are specified, one can solve the integrals in Eqs.~\eqref{int-chi_s} and~\eqref{int-h_i} and obtain the weak--field metric describing the surrounding spacetime of a slowly rotating source. Some explicit examples will be presented in Sec.~\ref{applic}.

It is useful to notice that the off--diagonal components $h_{i}$ of the metric do not depend on the function $f_0$ (see Eq.~\eqref{h_i diff}), which means that at linear order these terms are not affected by the form factor $\mathcal{F}_1$. From the physical perspective, this occurs because $\mathcal{F}_1$ modifies only the scalar part of the propagator~\eqref{Propagator}, which couples to the trace of the stress-energy tensor. Insomuch as the components $T_{0i}$ do not contribute to the trace $g^{\mu\nu}T_{\mu\nu}$  in the linearised regime, they do not act as sources for $h_{i}$. The situation is similar to what happens with the interaction between light and a static gravitational field~\cite{Accioly98,Accioly:2016qeb,Giacchini:2018twk}, where the scalar part of the propagator couples to the trace of the photon's stress--energy tensor---which is null.

\section{Light bending by a slowly rotating source}\label{se-light-bend}

In this Section we study the gravitational bending of light caused by a slowly rotating source whose surrounding spacetime geometry can be well described in terms of the metric~\eqref{polar-metric2}.  To compute the deflection angle we apply the relatively new method introduced by Gibbons and Werner~\cite{Gibbons:2008rj} based on the Gauss--Bonnet\footnote{The Gauss--Bonnet theorem is an important result of the differential geometry of surfaces, being  proved and discussed in most of the textbooks on the subject---see, \textit{e.g.},~\cite{Manfredo}.} theorem. Its original formulation assumed the spacetime to be spherically symmetric~\cite{Gibbons:2008rj}, and it was subsequently generalized to include stationary axisymmetric spacetimes and finite--distance corrections~\cite{Ono:2017pie,Werner:2012rc,Ishihara:2016vdc} (see, \textit{e.g.},~\cite{Ishihara:2016sfv,Jusufi:2018jof,Crisnejo:2018uyn,Jusufi:2018kry,Ono:2018ybw,Crisnejo:2018ppm,Javed:2019ynm,Crisnejo:2019ril,Jusufi:2019caq,Li:2020dln} and references therein for further developments and applications of the method). In our calculations we follow the scheme presented in~\cite{Ono:2017pie} and assume the limit of infinite distance between source and observer, so that the lens can be approximated as point--like. Also, for simplicity we restrict our considerations to motion on the equatorial plane.

The first step consists in observing that given a metric in the form
\begin{equation} \label{metric-AB}
{\rm d}s^2=- A(r)\, {\rm d}t^2 + 2 \zeta(r)\sin^2\theta\,{\rm d}\varphi\,{\rm d}t + B(r) ({\rm d}r^2+r^2{\rm d}\Omega^2)\,,
\end{equation}
such as~\eqref{polar-metric2}, one can define an auxiliary spatial metric $\ga_{ij}$ such that the null condition ${\rm d}s^2=0$ can be solved as~\cite{Ono:2017pie}
\ba
{\rm d}t = \sqrt{\gamma_{ij}{\rm d}x^i {\rm d}x^j} + \beta_i {\rm d}x^i ,
\ea
with
\ba
\gamma_{ij}{\rm d}x^i {\rm d}x^j = \frac{B(r)}{A(r)} \left(   {\rm d}r^2 +  r^2 \, {\rm d}\theta^2 \right)  + \left[\frac{B(r)}{A(r)}r^2 +  \frac{\zeta^2(r)\sin^2\theta}{A^2(r)} \right] \sin^2\theta \, {\rm d}\varphi^2
\ea
and
\ba
\beta_i {\rm d}x^i = \frac{\zeta(r)}{A(r)} \sin^2\theta \, {\rm d}\varphi .
\ea
In the absence of the off--diagonal term in the metric~\eqref{metric-AB}, \textit{i.e.}, if $\zeta(r) \equiv 0$, the spatial metric $\ga_{ij}$ would correspond to the optical metric and the trajectory of the light ray would be described as a geodesic with respect to it.

Because of the non--vanishing angular momentum of the source, the orbit of the light ray is no longer a geodesic on the Riemmanian space defined by $\ga_{ij}$. Indeed, in Ref.~\cite{Ono:2017pie} it was shown that, for motion on the equatorial plane $\theta=\pi/2$, the geodesic curvature along the light ray's path reads
\ba
\kappa_g(r) & = & - \sqrt{\frac{\gamma_{\theta\theta}}{\det \gamma_{ij}}} \frac{\partial \beta_\varphi}{\partial r},
\ea
which is clearly non--zero if $\beta_\varphi$ depends on $r$. 

In the explicit case of the metric~\eqref{polar-metric2} of a general linearised quadratic gravity theory, the three--dimensional auxiliary metric up to first order in $G$ is given by
\ba
\gamma_{rr} & = &  1 - 2 \chi_2(r)\,,
\\[3mm]
\gamma_{\theta\theta} & = & \left[  1 - 2 \chi_2(r) \right]r^2\,,
\\[3mm]
\gamma_{\varphi\varphi} & = &  \left[  1 - 2 \chi_2(r) \right]r^2 \sin^2\theta \,,
\ea
and
\ba
\beta_\varphi & = & \zeta(r) \sin^2\theta\, .
\ea
Hence, the geodesic curvature (along the light ray orbit on the equatorial plane) is
\ba 
\kappa_g(r) & = &  - \frac{1}{r} \frac{\partial \zeta(r)}{\partial r}\nonumber
\\[3mm]
&=& -\nonumber  \frac{2 G J}{r} \left[ \frac{\partial \mathcal{G}_2(r)}{\partial r} + r \frac{\partial^2 \mathcal{G}_2(r)}{\partial r^2} \right]
\\[3mm]
&=&  a \left[ \frac{1}{r} \frac{\partial \chi_2(r)}{\partial r} +  \frac{\partial^2 \chi_2(r)}{\partial r^2} \right]\,, \label{kappa_g-final}
\ea
where we only kept terms up to first order in $G$.

The second step of the method comprises the evaluation of the Gaussian curvature of the surface parametrized by $(r,\varphi)$ with the restriction of the metric ${\ga}_{ij}$ to the subspace $\theta=\pi/2$. This gives~\cite{Ono:2017pie}
\ba
K(r) & = & - \frac{1}{2} \sqrt{\frac{A^3}{B(ABr^2+\zeta^2)}} \frac{\partial}{\partial r} \left[ \sqrt{\frac{A^3}{B(ABr^2+\zeta^2)}} \frac{\partial}{\partial r} \left( \frac{ABr^2+ \zeta^2}{A^2} \right) \right],
\ea
where for economy of notation we omitted the dependence of $A$, $B$ and $\zeta$ on the coordinate $r$. For the linearised metric~\eqref{polar-metric2} one has, explicitly,
\ba
K(r) =  \frac{1}{r} \frac{\partial \chi_2(r)}{\partial r} +  \frac{\partial^2 \chi_2 (r)}{\partial r^2}\,.
\label{curvature}
\ea
Comparing the formulas for the Gaussian and the geodesic curvature~\eqref{kappa_g-final} it happens that 
\ba
\kappa_g(r) =  a K(r),
\ea
where we recall that $a=J/M$ is the rotational parameter.

\begin{figure}[t]
	\begin{center}
		\includegraphics[scale=0.55]{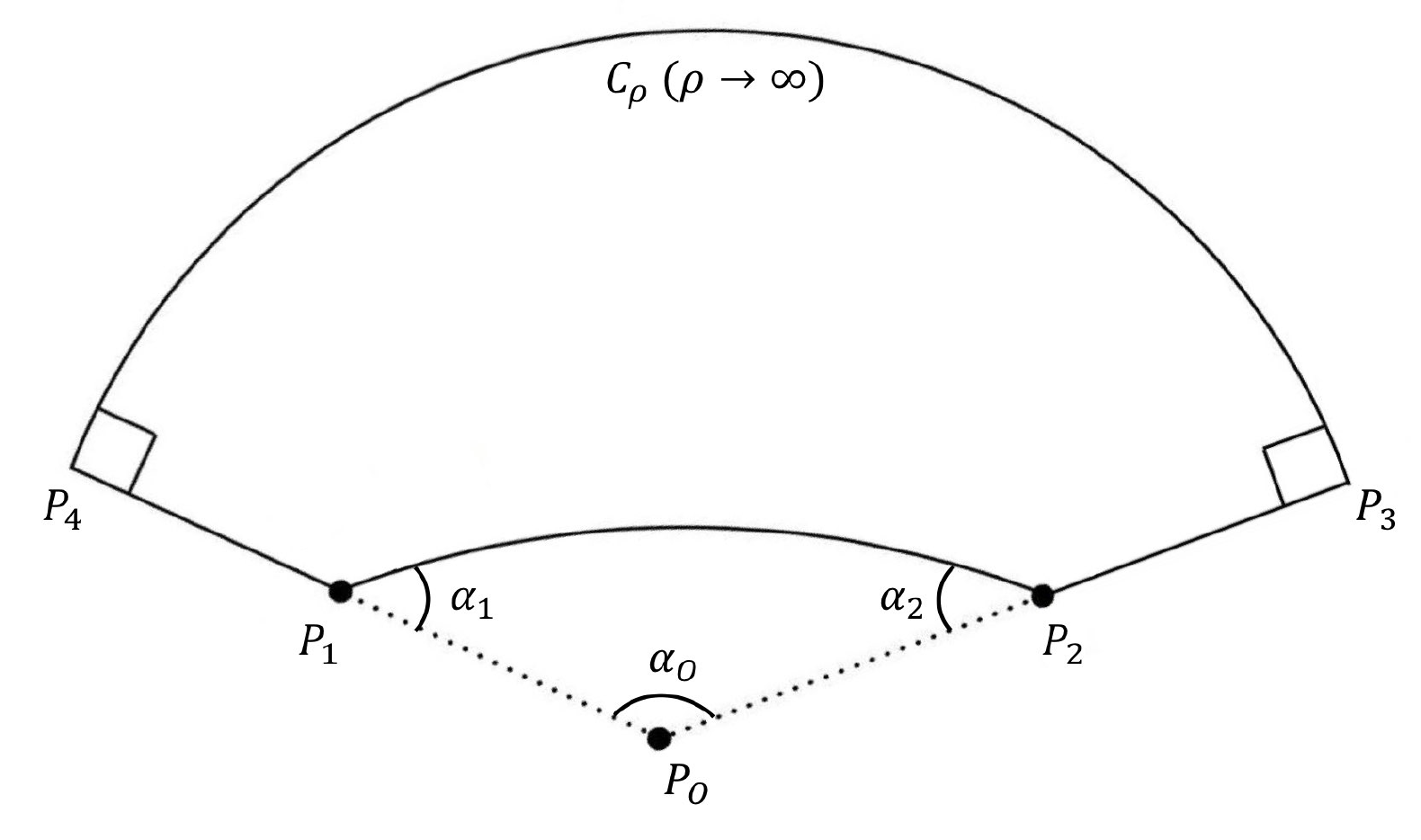}
		\caption{\footnotesize \label{Fig1} Illustration of the relevant domain for the application of the Gauss--Bonnet theorem in order to evaluate the deflection angle $\alpha$ undergone by a light ray emitted in $P_1$ and observed in $P_2$. The quadrilateral $\mathscr{D}$ corresponds to the region defined by $P_1P_2P_3P_4$, the light ray's path is $\overline{P_1P_2}$ and $C_\rho$ is an arc of circle with radius $\rho$. The point $P_O$ is the origin of the coordinate system and the centre of the mass distribution. The deflection angle $\alpha$ can be defined from the triangle $P_1P_2P_O$, whose internal angles sum to $\alpha_1+\alpha_2+\alpha_O=\pi + \alpha$. By using this relation and applying the Gauss--Bonnet theorem on $\mathscr{D}$ taking the limit $\rho \rightarrow \infty$ it is possible to derive the coordinate--invariant expression~\eqref{defl-angle}~\cite{Ono:2017pie}.}
	\end{center}
\end{figure}

With the expressions for the Gaussian curvature and the geodesic curvature along the trajectory of the light ray, one can use the Gauss--Bonnet theorem to evaluate the deflection angle $\alpha$ between a source $P_1$ and an observer $P_2$. In fact, applying the theorem to the domain $\mathscr{D}$ depicted in Fig.~\ref{Fig1}, it follows~\cite{Ono:2017pie} 
\ba 
\alpha =  -  \iint_{\mathscr{D}} K \, {\rm d}S + \int_{P_1}^{P_2} \kappa_g \, {\rm d}\ell  .\label{defl-angle}
\ea

We note that this method is very general and allows the calculation of finite--distance corrections to the bending of light~\cite{Ono:2017pie,Ishihara:2016vdc}. For simplicity, however, we shall only consider the case in which source and receiver are at infinity. In this case $\mathscr{D}$ is defined by the range of angles $\varphi\in(-\pi/2,+\pi/2)$, while the coordinate $r$ is bounded from below by the light ray orbit.
In the linear approximation one can parametrize the trajectory by $r_\text{orb}(\varphi) = b / \cos \varphi$ and $\ell = b \vert \tan \varphi \vert$~\cite{Ono:2017pie}, where $b$ is the impact parameter and $\varphi$ is measured such that 
the closest approach point  corresponds to $\varphi=0$ 
(while the source/observer are located at $\varphi = \mp \pi/2$).
Then, given the  surface element
\begin{equation}
{\rm d}S=\sqrt{\gamma_{rr}\gamma_{\varphi\varphi}}\,{\rm d}r{\rm d}\varphi=r {\rm d}r{\rm d}\varphi,
\end{equation}
we obtain 
\ba \label{I-1}
I_1 \, \equiv \, - \iint_{\mathscr{D}} K \, {\rm d}S \,=\,   - 2 \int_{0}^{\pi/2}  \int_{0}^{\frac{\sin \varphi}{b}} \frac{K(1/u)}{u^3}  \, {\rm d}u{\rm d}\varphi \,,
\ea
where we have employed the change of variable $u \equiv 1/r$ and kept terms only up to linear order in~$G$. Since in the approximation we consider $K$ does not depend on $\zeta$, it turns out that~\eqref{I-1} should coincide with the deflection angle in the case of a static metric.

As for the integral of the geodesic curvature along the light ray path, one should note that the sign of the distance element ${\rm d}\ell$ depends on whether the motion is prograde or retrograde. Therefore, in the linear approximation we set ${\rm d}\ell = \sigma b \sec^2\varphi \, {\rm d}\varphi$, where $\sigma = +1$ for the prograde motion, and $\sigma = -1$ for retrograde motion. This yields
\ba \label{I-2}
I_2 \, \equiv \, \int_{P_1}^{P_2} \kappa_g \, {\rm d}\ell \, =  \, 2 \sigma b \int_{0}^{\pi/2} \frac{\kappa_g( b / \cos \varphi)}{\cos^2 \varphi} \, {\rm d}\varphi \, .
\ea

Putting together the contributions $I_1$ and $I_2$ we have the expression for the deflection angle in Eq.~\eqref{defl-angle}:
\begin{eqnarray}
\alpha &=& - 2 \int_{0}^{\pi/2} \int_{0}^{\frac{\sin \varphi}{b}} \frac{K(1/u)}{u^3}  \, {\rm d}u  {\rm d}\varphi 
+
 2 \sigma b \int_{0}^{\pi/2} \frac{\kappa_g( b / \cos\varphi)}{\cos^2 \varphi} \, {\rm d}\varphi\, + \mathcal{O}(G^2)\,.\label{complete deflec angle}
\end{eqnarray}

We remark that at linear order both the Gaussian and the geodesic curvatures only depend on the potential $\chi_2(r)$, which means that the modifications on the scalar sector of the theory (via a function $f_0 \neq 1$) do not affect the trajectory of a light ray. This extends the result of Ref.~\cite{Giacchini:2018twk}, which was restricted to the case of static spherically symmetric metrics.

\section{Application to several gravitational theories}\label{applic}

We now apply the formalism presented in the previous Section to some of the most popular quadratic theories of gravity, described by different choices of form factors, and for each of them we evaluate the expression for the deflection angle. Before doing that, it is instructive to show the calculation for the case of Einstein's GR, as this result should be matched by the extended theories in the appropriate limits.

General relativity is recovered when the form factors in the action~\eqref{quad-action} are zero, so that the relevant metric potentials are given by
\begin{equation} \label{PotGR}
\chi_2^{\text{GR}}(r) = -\frac{2 GM}{r}\,,\qquad \zeta^{\text{GR}}(r)=-\frac{2GMa}{r}\,; 
\end{equation}
while the geodesic and the Gaussian curvatures read~\cite{Ono:2017pie}
\ba
\kappa_g(r) = aK(r)=- \frac{2 a G M}{r^3}\,.
\ea
Then, the static and rotational contributions to the deflection angle are (see~\eqref{I-1} and~\eqref{I-2})
\ba
I_1  =  \frac{4 G M}{b} \,, \qquad I_2 = - \sigma \frac{4 a G M}{b^2}\,,
\ea
which yield
\ba
\alpha_{\text{GR}} = \frac{4 G M}{b} - \sigma \frac{4 a G M}{b^2}\,. \label{defl-ang GR}
\ea
The first term is the standard bending angle for a static source in the point--like approximation, whereas the second term is the contribution associated to the source's rotation~\cite{Plebanski,Boyer-Lindquist}. Notice that while the former is of order $M b^{-1}$, the latter is proportional to $a M b^{-2}$, which is usually much smaller~\cite{Edery:2006hm} (for example, $a \approx 0.3 \text{ km}$ for the Sun~\cite{Pijpers-Bi}).

\subsection{$\mathcal{R}\mathcal{F}(\Box)\mathcal{R}$--gravity}\label{SecFR}

Let us now analyse an extension of GR in which only the scalar part of the propagator is modified,
\textit{i.e.}, 
\begin{equation} \label{fR}
\mathcal{S}=\frac{1}{2\kappa^2}\int {\rm d}^4x\sqrt{-g} \left[  \mathcal{R}+\frac{1}{2}  \mathcal{R}\mathcal{F}_1(\Box)\mathcal{R}  \right] \,,
\end{equation}
where $\mathcal{F}_1$ is an arbitrary form factor, and $\mathcal{F}_2=0$ which implies $f_2\equiv 1$ just like in GR. It is clear, thus, that the spin--$2$ field $\chi_2$ and the cross--term $\zeta$ are given by~\eqref{PotGR}. Therefore, the expression for the deflection angle is still given by Eq.~\eqref{defl-ang GR}, as the modifications in the scalar part of the theory do not  affect the path followed by the light ray.

As mentioned in the end of Section~\ref{se-light-bend}, this generalizes the result of~\cite{Giacchini:2018twk} to the case of slowly rotating linearised metrics and it happens because the off--diagonal components do not depend on the spin--$0$ sector (see Eq.~\eqref{h_i diff}). This fact can be also explained by noticing that the metric $g_{\mu\nu}^{\text{ext}}$ given by~\eqref{polar-metric2} for the extended theory\footnote{That is, the potentials of the metric $g_{\mu\nu}^{\text{ext}}$ are $\chi_2= \chi_2^{\text{GR}}$, $\zeta=\zeta^{\text{GR}}$ and $\chi_0$ is arbitrary (depending on~$\mathcal{F}_1$).}~\eqref{fR} is conformally related to the corresponding metric in GR. Indeed, 
\begin{equation} 
g_{\mu\nu}^{\text{ext}} =  \left[ 1 + \frac{2}{3} \left( \chi_0 - \chi_0^{\text{GR}} \right)  \right] g_{\mu\nu}^{\text{GR}} + \mathcal{O}(G^2),
\end{equation}
where $g_{\mu\nu}^{\text{GR}}$ is the metric~\eqref{polar-metric2} for GR, \textit{i.e.}, with the potentials $\chi_0^{\text{GR}} = - \chi_2^{\text{GR}}/2$ and $\zeta^{\text{GR}}$ of Eq.~\eqref{PotGR}.
Inasmuch as null geodesics are invariant under conformal transformations, it follows that in the linear regime of the theory described by~\eqref{fR} the trajectory of a light ray is the same for any of the metrics $g_{\mu\nu}^{\text{ext}}$ defined by an arbitrary potential $\chi_0$. 

Further discussion concerning the deflection of light in particular models of the type~\eqref{fR} can be found in Refs.~\cite{Capozziello06,Clifton,Capozziello11,Berry-Gair,Stabile&Stabile,LWLM,Accioly98}. In particular, we remark that even though in the linear approximation the path of the light ray is the same in this class of models, it is possible to distinguish between them by using the light bending in combination with other measurements~\cite{Clifton,Berry-Gair,Accioly:2016etf,Giacchini:2018twk}; we shall return to this point in Section~\ref{conclus}.

\subsection{Fourth--derivative gravity}\label{Sec4Der}

Let us now consider Stelle's fourth--derivative gravity~\cite{-K.-S.}, which corresponds to
\begin{equation}
\mathcal{F}_1=c_1 \,,\qquad\mathcal{F}_2=-c_2\qquad\Rightarrow\qquad f_2=1-\frac{c_2}{2}\,\Box\,,
\end{equation}
where $c_1$ and $c_2$ are positive constants.
Unlike the previous case, we now have a contribution coming from $\mathcal{F}_2$ which affects the spin--$2$ field, thus modifying the deflection angle non--trivially. Indeed, we have
\ba
\chi_2(r) = -\frac{2GM}{r} \left( 1 - e^{-m_2 r} \right)\,,\qquad \zeta(r)=-\frac{2GMa}{r}\left[1-(1+m_2r)e^{-m_2r}\right]\,,
\ea
being $m_2=\sqrt{2/c_2}$ the mass of the massive spin--$2$ component. The geodesic and Gaussian curvatures are obtained by substituting the expression above in~\eqref{kappa_g-final} and~\eqref{curvature}:
\ba \label{Curv4th}
\kappa_g(r)=aK(r) = - \frac{2 a G M}{r^3} \left[ 1 - e^{-m_2 r} - m_2 r e^{-m_2 r} (1 + m_2 r) \right]\,.
\ea
Their contribution to the deflection angle read
\ba
I_1 & = & \frac{4 G M}{b} \left[ 1 -  \int_0^{\pi/2} e^{-m_2 b \csc\varphi} \left( b m_2 + \sin\varphi \right) \, {\rm d}\varphi \right] 
\ea
and
\ba
I_2& = & - \sigma \frac{4 a G M}{b^2} \left[ 1 -  \int_0^{\pi/2} e^{-m_2 b \sec \varphi} \left( \cos\varphi + bm_2 + b^2 m_2^2 \sec\varphi \right) \, {\rm d}\varphi \right] \,.
\ea
The effect of the repulsive force of the massive spin--$2$ ghost~\cite{Accioly:2015fka,Newton-MNS} is manifest in the previous equations. In fact, all the terms which depend on $m_2$ appear with opposite sign with respect to the terms of GR, contributing to make the curvatures smaller (in absolute value).

As mentioned before, the term $I_1$ corresponds to the static contribution to the deflection angle. Notice that through the change of variable in the form $b \csc\varphi = \sqrt{b^2 + x^2} \equiv \rho(x)$, it can be written as
\ba
I_1 = \frac{4 G M}{b} - 4 G M b \int_0^\infty \frac{e^{-m_2 \rho(x)}}{\rho^2(x)} \left[ m_2 + \frac{1}{\rho(x)} \right] \, {\rm d}x.
\ea
This expression matches the results obtained by means of other techniques in Refs.~\cite{Accioly:2015fka,Accioly98,Accioly:2016qeb} and verifies the consistency of our calculations using the Gauss--Bonnet theorem.

Making the further change of integration variable,  $y=b/\sqrt{b^2+x^2},$ the final expression for the deflection angle in the fourth--derivative gravity can be recast in the following more compact form:
\ba
\alpha = \alpha_{\text{GR}}  - \frac{4 G M}{b} \int_0^{1} \frac{e^{-b m_2 / y}}{\sqrt{1-y^2}} \left[ \left( 1 - \frac{\sigma a}{b} \right)  \left( b m_2 + y \right) - \frac{\sigma a b m_2^2}{y}  \right]  \, {\rm d}y .
\ea

\subsection{Sixth--derivative gravity with complex poles}\label{Sec6Der}

Another model for which the deflection angle is known in the case of a static weak--field metric is the sixth--order gravity~\cite{Accioly:2016qeb}. This model can be super--renormalizable~\cite{Asorey:1996hz} and it is the most simple one that admits complex poles in the propagator, in an attempt to conciliate unitarity and renormalizability in perturbative quantum gravity~\cite{ModestoShapiro:2015ozb,Modesto:2015ozb}. There are three possible scenarios for this theory, depending on whether the poles are real or complex, simple or degenerate. Here we only show the explicit calculation for the most interesting case of a pair of complex poles (also known as Lee--Wick gravity); moreover, for simplicity, we assume that the real and the imaginary parts of the poles are equal. The result for a general polynomial--derivative theory, from which the omitted cases can be easily deduced, is presented in the next section.

The model under consideration is defined by ($c_1,c_2>0$)
\begin{equation}
\mathcal{F}_1=c_1\,\Box\,,\qquad \mathcal{F}_2=-c_2\,\Box \qquad \Rightarrow\qquad f_2=1-\frac{c_2}{2}\,\Box^2\,,
\end{equation}
which yield the potentials~\cite{Modesto:2015ozb,Accioly:2016qeb}
\ba
\chi_2(r) &=& -\frac{2GM}{r} \left( 1 - e^{-m_2 r}{\rm cos}\,m_2r \right)\,,\\[3mm] \zeta(r)&=&-\frac{2GMa}{r}\left[1-e^{-m_2r}(1+m_2r)\,{\rm cos}\,m_2r-e^{-m_2r}m_2r\,{\rm sin}\,m_2r\right]\,,
\ea
where, $m_2=\sqrt{2/c_2}$ as in the previous example. The oscillatory contribution, which turns out to be damped by a Yukawa potential, is typical of models with complex poles in the propagator~\cite{Giacchini:2016xns}. For the geodesic and Gaussian curvatures we obtain
\be \label{Curv6th}
\kappa_g(r)= aK(r)= - \frac{2 a G M}{r^3} \left\lbrace  1 - e^{-m_2 r} \left[ (1+m_2 r){\rm cos}\,m_2 r + m_2 r\,(1 + 2m_2 r){\rm sin}\,m_2 r \right] \right\rbrace ,
\ee
from which it follows
\be \label{6derI1}
I_1 = \frac{4 G M}{b} \left\lbrace  1 - \int_0^{\pi/2}  e^{-m_2 b \csc\varphi} 
\left[   \left( b m_2 + \sin\varphi \right) \cos(m_2 b \csc\varphi) + m_2 b \sin(m_2 b \csc\varphi) \right]    {\rm d}\varphi \right\rbrace  \,
\ee
and
\ba
I_2 &=& - \sigma \frac{4 a G M}{b^2} \bigg\lbrace   1 -  \int_0^{\pi/2} e^{-m_2 b \sec \varphi} \left[ \cos\varphi\, (1+m_2 b\,\sec\varphi)\,\cos(bm_2\sec\varphi)\right.\nonumber
\\[3mm]
&&\qquad\qquad\qquad\qquad\qquad\left.+ \, bm_2\,(1+ 2b m_2 \sec\varphi)\,{\rm sin}(bm_2\sec\varphi) \right]\, {\rm d}\varphi  \bigg\rbrace  \,.
\ea

Again, expression~\eqref{6derI1} should be compared to the bending angle for a static weak--field evaluated in~\cite{Accioly:2016qeb} by means of the optical--mechanical analogy. Both results agree after one implements the change of variable in the form $b \csc\varphi = \sqrt{b^2 + x^2} \equiv \rho(x)$.
As done in the case of fourth--order gravity, we can make another change of integration variable and cast the deflection angle in the more compact form
\begin{eqnarray}
\alpha&=& \alpha_{\text{GR}}  - \frac{4 G M}{b}  \int_0^{1}  \frac{e^{-m_2 b/y}}{\sqrt{1-y^2}} 
\left\lbrace  \left(1-\frac{\sigma a}{b}\right)  \left( b m_2 + y \right) \cos(m_2 b/y)\right. 
\nonumber
\\[3mm]
&&\qquad\quad\left.+ m_2  \sin(m_2 b/y)\left[1-\frac{\sigma a}{b}\left(1+\frac{2m_2b}{y}\right)\right] \right\rbrace   {\rm d}y \,.
\end{eqnarray}

\subsection{General polynomial--derivative gravity}\label{App}

The two previous examples are particular cases of polynomial--derivative gravity, which is defined by real polynomial form factors $\mathcal{F}_{1,2}$. The function $f_{2}$ is then a polynomial too, being factored as
\ba
f_2(z) = \prod_{i=1}^n \left( \frac{m_i^2 - z}{ m_i^{2}}\right) ^{\alpha_i} ,
\ea
where $z=m_i^2$ is one of the $n$ roots of the equation $f_2(z)=0$, $\alpha_i$ is its multiplicity and $N=\sum_{i} \alpha_i$ is the degree of $f_2(z)$. Furthermore, we assume that $\text{Re}(m_i) >0$~\cite{Accioly:2016qeb,Giacchini:2018gxp}. As a consequence, the potentials read~\cite{Giacchini:2018gxp}
\ba \label{Chi2Poly}
\chi_2(r) = -\frac{2GM}{r} + \frac{2GM}{\sqrt{\pi}} \sum_{i=1}^n \sum_{j=1}^{\alpha_i} c_{i,j} \left( \frac{r}{2 m_i} \right)^{j-\frac{3}{2}} K_{j-\frac{3}{2}} (m_i r ) \, ,
\ea
\ba
\zeta(r) = -\frac{2GMa}{r} + \frac{2GMar}{\sqrt{\pi}} \sum_{i=1}^n \sum_{j=1}^{\alpha_i}  m_i \, c_{i,j}\left( \frac{r}{2 m_i} \right)^{j-\frac{3}{2}} K_{j-\frac{5}{2}} (m_i r ) \, ,
\ea
where $c_{i,j}$ are coefficients given by combinations of the mass parameters $m_i$,
%
\ba
\label{cij}
c_{i,j} = \frac{-1}{(\alpha_i - j)! (j - 1)!} \, \frac{{\rm d}^{\alpha_i -j}}{{\rm d}z^{\alpha_i-j}} \frac{\left( z + m_i^2\right)^{\alpha_i}}{zf_2(-z)}  \Bigg|_{z = - m_i^2} \, ,
\ea
and $K_{\nu}$ is the modified Bessel function of the second kind.

The corresponding geodesic and Gaussian curvatures read
\ba \label{CurvGenPol}
\kappa_g(r) = aK(r) &=& - \frac{2 a G M}{r^3} + \frac{4 a G M}{r^4\sqrt{\pi}} \sum_{i=1}^n \sum_{j=1}^{\alpha_i} c_{i,j}\left( \frac{r}{2 m_i} \right)^{j-\frac{1}{2}} 
\nonumber
\\
&& \times \, \bigl\{ \left[ - (3 - 2j)^2 (2j - 1) - 4 ( j - 1) m_i^2 r^2 \right] K_{j-\frac{1}{2}}(m_i r ) 
\nonumber
\\
&&\,\,\,\,\,\,\,\,\, + \, m_i r \left[ (2j - 3)^2 + m_i^2 r^2 \right] K_{j+\frac{1}{2}}(m_i r ) \bigl\}
\,.
\ea
It is possible to carry on the computation of the formula for the deflection angle, which is expressed by integrals involving generalized hypergeometrical functions. We omit these cumbersome expressions which can be easily calculated for the particular model of interest by inserting~\eqref{CurvGenPol} into~\eqref{complete deflec angle}.

\subsection{Analytic nonlocal gravity}
\label{SANLG}

We now consider one example of gravitational theory whose action contains non--polynomial differential operators. For the moment we shall consider only the case of analytic operators, postponing the analysis of two models with non--analytic operators to the next Subsection. Namely, here we assume that the function $f_2$ is the exponential of an entire function. The main virtue of these functions is the absence of unhealthy massive poles ({\it i.e.,} ghosts) in the propagator and the possibility to have a (super--)renormalizable theory of gravity. A variety of such models has been considered in the literature~\cite{Krasnikov:1987yj,Kuzmin:1989sp,Tomboulis:1997gg,Biswas:2005qr,Modesto:2011kw,Biswas:2011ar,Buoninfante:2018bkc,Buoninfante:2019der,Buoninfante:2019uwo,Giacchini:2018wlf,Biswas:2013cha,Biswas:2016etb,Edholm:2016hbt,Buoninfante:2018xiw,Buoninfante:2018rlq,Buoninfante:2018stt,Buoninfante:2018mre,Buoninfante:2018xif,Buoninfante:2018gce,Buoninfante:2018lnh,Buoninfante:2019swn,Buoninfante:2019teo,Abel:2019zou,Buoninfante:2020ctr,Buoninfante:2017kgj,Buoninfante:2017rbw,SravanKumar:2019eqt,Koshelev:2017tvv,Koshelev:2020foq,BFZ-pbranes,Calcagni-Universe,Ercan18,Kilicarslan:2019njc,Li:2015bqa,CMNa,Calcagni:2018pro,Briscese:2019rii,Jens,Boos:2020kgj,Boos:2020ccj,Frolov:Exp,Frolov:Poly}. For simplicity here we only analyse one of the simplest choices, given by 
\begin{equation}
\mathcal{F}_1=-\frac{1}{2}\mathcal{F}_2=\,\frac{1-e^{-\Box/\mu^2}}{2\,\Box}\,\qquad\Rightarrow\qquad f_2=e^{-\Box/\mu^2}, \label{ghost--free-choice}
\end{equation}
where $\mu$ is the new energy scale at which nonlocal effects should manifest.

For this theory, the field potentials are
\ba
\chi_2(r) = -\frac{2 G M}{r} \text{Erf} \left( \frac{\mu r}{2} \right)\,,\qquad \zeta(r)=-\frac{2GMa}{r}\left[{\rm Erf}\left(\frac{\mu r}{2}\right)-\frac{\mu r\,e^{-\frac{\mu^2 r^2}{4}}}{\sqrt{\pi}}\right]\,,
\ea
from which we obtain the curvatures
\ba
\kappa_g(r) = a K(r) & = & - \frac{2 a G M}{r^3} \left[  \text{Erf}  \left( \frac{\mu r}{2} \right) -  \left( 1+\frac{\mu^2r^2}{2}\right) \frac{e^{-\frac{\mu^2r^2}{4}}\mu r }{\sqrt{\pi}} \right].
\ea
Thanks to the presence of Gaussian functions, in this case all the integrals can be performed analytically, yielding a compact and elegant form for the deflection angle. Indeed, we have
\ba
I_1  = \dfrac{4 G M}{b} \left( 1 - e^{-\frac{\mu^2 b^2}{4}} \right)\,
\ea
and
\ba
I_2  = - \sigma \dfrac{4 a G M}{b^2} + \sigma \dfrac{2 a G M}{b^2}\left( 2 + \mu^2 b^2 \right) e^{-\frac{\mu^2 b^2}{4}}\,,
\ea
which give
\ba
\alpha = \alpha_{\text{GR}}  - \dfrac{4 G M}{b} e^{-\frac{\mu^2 b^2}{4}} \left[ 1 - \sigma \, \frac{a}{b} \left( 1 + \frac{\mu^2 b^2}{2} \right)  \right]   \,.
\ea
Notice that in the limit $\mu\rightarrow\infty$ the previous expression consistently reproduces the deflection angle in GR~\eqref{defl-ang GR}.

It is also worth to investigate the opposite limit, {\it i.e.,} $\mu b\ll 1,$ in which the gravitational interaction is highly nonlocal:
\ba
\alpha = GM\mu^2b \left(1+\frac{\sigma a}{b}\right)  +\mathcal{O}(\mu^2 b^2) \,.\label{nonlocal-lim}
\ea
Note that at zeroth order in $\mu b$ the deflection angle vanishes as the GR piece is compensated by an equal and opposite term, while it starts acquiring a non--vanishing purely nonlocal contribution at order $\mathcal{O}(\mu b).$ In this regime the impact parameter $b$ is engulfed by the nonlocal length scale $\mu^{-1}$. 

The result $\lim_{\mu b \rightarrow 0} \alpha = 0$ is a consequence of the suppression of gravity in the limit of small distances, and it holds for all the quadratic gravity theories which have a bounded potential $\chi_2$. In this more general case, it happens when $b$ is much smaller than the other length scales of the model. For example, for the polynomial--derivative theory considered in Section~\ref{App} it occurs for $m b \ll 1$, where $m = \max_i\left\lbrace \text{Re} (m_i) \right\rbrace$. This can be verified in a straightforward manner by noticing that the Gaussian and geodesic curvatures~\eqref{Curv4th} and~\eqref{Curv6th} tend to zero  in the limit $m_2 \rightarrow 0$, and  the same can be shown for the more general model in Eq.~\eqref{CurvGenPol} by applying the results of the works~\cite{Giacchini:2016xns,Giacchini:2018wlf} (see also~\cite{Giacchini:2018gxp,Newton-MNS}).

\subsection{Non--analytic nonlocal gravity} \label{SecNANLG}

As last examples, we consider two nonlocal models whose gravitational actions are constructed in terms of non--analytic differential operators, which lead to infrared extensions of Einstein's GR. Such kind of nonlocal terms can be introduced as an effective treatment of quantum corrections to the gravitational action and reproduce the renormalization group running of the cosmological constant and the Einstein--Hilbert term~\cite{GoSh,Deser:2007jk,Maggiore2}.

\subsubsection{First model: $\Box^{-1}$}

The first nonlocal action that we study is an extension of the model proposed by Deser and Woodard~\cite{Deser:2007jk}, and it is characterized by the following form factors~\cite{Ferreira:2013tqn}:
\begin{equation}
\mathcal{F}_1=\frac{c_1}{\Box}\,,\qquad \mathcal{F}_2=\frac{c_2}{\Box}\,\qquad\Rightarrow\,\qquad f_2=1+\frac{c_2}{2}\,,\label{nonlocal-choice1}
\end{equation}
and the two relevant field potentials read~\cite{Conroy:2014eja}
\begin{equation}
\chi_2(r) = -\frac{1}{1+c_2/2}\frac{2 GM}{r}\,,\qquad \zeta(r)=-\frac{1}{1+c_2/2}\frac{2GMa}{r}\,.
\end{equation}
The geodesic and the Gaussian curvature are similar to those of GR, but with the rescaling factor $(1+c_2/2)^{-1}$,
\ba
\kappa_g(r) =aK(r) =- \frac{1}{1+c_2/2}\frac{2 a G M}{r^3}\,.
\ea
Then, it follows
\ba
I_1  =  \frac{4 G M}{b(1+c_2/2)} \,,\qquad I_2 = - \sigma \frac{4 a G M}{b^2(1+c_2/2)},
\ea
so that the deflection angle is
\ba
\alpha =\frac{\alpha_{\text{GR}}}{1+c_2/2} \,,
\ea
which recovers GR in the limit $c_2\rightarrow 0$, as expected.

\subsubsection{Second model: $\Box^{-2}$}

Let us now choose the following form factors~\cite{Maggiore2,Maggiore1,Cusin:2015rex,Kumar:2018pkb} ($c_1,c_2>0$)
\begin{equation}
\mathcal{F}_1=\frac{c_1}{\Box^2}\,,\qquad\mathcal{F}_2=-\frac{2 c_2}{\Box^2}\,\qquad\Rightarrow\,\qquad f_2=1-\frac{c_2}{\Box}\,.\label{nonlocal-choice2}
\end{equation}
In this case we do not get a simple constant factor as modification, but Yukawa potentials~\cite{Conroy:2014eja}: 
\ba
\chi_2(r) = -\frac{2GM}{r}e^{-\mu_2 r}\,,\qquad \zeta(r)=-\frac{2GMa}{r}(1+\mu_2r)e^{-\mu_2r}\,,
\ea
where the mass of spin--$2$ component is now given by $\mu_2=\sqrt{c_2}$.
The Newtonian potential is, thus, screened by this massive parameter in such a way that the usual form proportional to $1/r$ is only observed for $r \mu_2 \ll 1$.

The geodesic and Gaussian curvatures are given by
\ba \label{Curv-box-2}
\kappa_g(r)=aK(r) = - \frac{2 a G M}{r^3} e^{-\mu_2 r}\left[ 1+ \mu_2 r(1+\mu_2 r) \right]\,.
\ea
Then, it follows
\ba
I_1 & = & \frac{4 G M}{b}\int_0^{\pi/2}e^{-\mu_2b\,\csc\varphi}(\mu_2b+\sin\varphi) \, {\rm d}\varphi
\ea
and
\ba
I_2 &=& \nonumber - \sigma \frac{4 a G M}{b^2}\int_0^{\pi/2}{\rm d}\varphi e^{-\mu_2b\, \sec \varphi}\left[\cos\varphi -  \mu_2 b\left( 1+ \mu_2 b \sec\varphi\right)\right]  \, {\rm d}\varphi\,.
\ea
By making the same changes of integration variables performed in the case of fourth-- and sixth--order gravity, we can obtain the following expression for the deflection angle:
\begin{eqnarray}
\alpha&=&\frac{4GM}{b} \int_0^1 \frac{e^{-\mu_2b/y}}{\sqrt{1-y^2}}\left\lbrace \mu_2 b+y
+\frac{\sigma a}{b}\left[y-\mu_2b\left(1-\frac{\mu_2b}{y}\right)\right]\right\rbrace \,{\rm d}y   \,.
\end{eqnarray}

Notice that since the Newtonian potential gets screened for $r \gtrsim \mu_2^{-1}$, it turns out that if the impact parameter $b$ is much larger than $\mu_2^{-1}$, all the trajectory of the light ray would be in a region of very small curvature (see Eq.~\eqref{Curv-box-2}), whence $\alpha \approx 0$. On the other hand, if $b \lesssim \mu_2^{-1}$ then $\alpha < \alpha_\text{GR}$, as part of the trajectory would be in a screened zone. Therefore, for the application of this model to particular systems it may be necessary to take into account finite--distance corrections. For example, if $\mu_2^{-1}$ is so large that the source and the observer are deep inside the potential, the deflection angle would be roughly the same as in GR. 

It is worthwhile emphasizing that this behaviour is opposite to what happens in the case of analytic nonlocal gravity analysed in Section~\ref{SANLG} (or, more generally, in higher-derivative gravity models). Indeed, in that case we had short--distance (ultraviolet) extensions of Einstein's GR and  deep inside the scale of nonlocality $\mu^{-1}$ the deflection angle was entirely controlled by the nonlocality of the gravitational interaction (see Eq.~\eqref{nonlocal-lim} and the subsequent discussion).


\section{Discussion and conclusions}\label{conclus}

The gravitational deflection of light was one of the first predictions of GR to be verified experimentally and remains among the classical tests of gravity models. In this work we presented a general scheme for evaluating the bending angle owed to a slowly rotating source in the context of linearised quadratic theories of gravity.  These kind of models can be viewed as extensions of GR in which the propagator of the gravitational interaction is modified in the scalar and/or spin-$2$ sectors. Each of the modifications has a different effect on the light bending~\cite{Accioly:2016etf,Accioly:2015fka,Accioly98,Accioly:2016qeb,Giacchini:2018twk}.

As a first application to theories beyond Einstein's GR, we considered gravitational models described by the action in Eq.~\eqref{fR} which entails only modification of the scalar component of the propagator. In Ref.~\cite{Giacchini:2018twk} it was shown that, for such models and for static spherically symmetric configurations, a modified spin--$0$ component does not play any active role in the interaction with light, in the sense that light rays' trajectories are unaffected. Here we extended this result to the more general case of slowly rotating metrics. Indeed, in Section~\ref{SecFR} it was explicitly shown that in the class of theories~\eqref{fR} light follows the same path as it does in GR, as $\chi_2 = \chi_2^{\text{GR}}$; therefore, $\alpha=\alpha_{\text{GR}}$. This does not mean, however, that light deflection cannot be used to discriminate between models of this type. In fact, as discussed in detail in~\cite{Giacchini:2018twk}, in order to predict the bending angle it is necessary to know the mass of the body which causes the deflection. This quantity is usually a Keplerian mass, determined by the investigation of orbits of massive bodies, and, therefore, it is model--dependent (as the interaction between non--relativistic objects depends on the scalar part of the potential, $\chi_0$). Taking this into account, in the static case it is possible to write the \textit{predicted} deflection angle $\hat{\alpha}$ in the form
\begin{equation} \label{hatAlfa}
\hat{\alpha} = \frac{1+\gamma}{2} \alpha_{\text{GR}}
\end{equation}
where the quantity
\begin{equation}
\gamma(\bar{r}) = \frac{\Psi(\bar{r})}{\Phi(\bar{r})} = \frac{\chi_2^{\text{GR}}(\bar{r}) - \chi_0(\bar{r})}{2 \chi_2^{\text{GR}}(\bar{r}) + \chi_0(\bar{r})}
\end{equation} 
is not a true constant, but depends on the scale $\bar{r}$ at which the (Keplerian) mass of the central body was determined. Of course, this assumes that at scales near $\bar{r}$ the potentials $\Phi$ and $\Psi$ can be sufficiently well approximated as being proportional to $1/r$. In a more general scenario, it would be possible to observe deviations from the Keplerian orbits, which could offer a more direct measurement of the mass $M$.

Furthermore, we considered extended gravity models in which  the spin--$2$ component of the propagator is modified, so that the potential $\chi_2 \neq \chi_2^{\text{GR}}$ plays an active and crucial role in the interaction with light~\cite{Giacchini:2018twk}, modifying the trajectory of light rays as we explicitly showed in the examples of Sections~\ref{Sec4Der}--\ref{SecNANLG}. In fact, the potential $\chi_2$ affects not only the static contribution to the deflection angle (which was already known after, \textit{e.g.},~\cite{Accioly98,Accioly:2016qeb,Giacchini:2018twk}) but also the terms which depend on the rotation of the source---see Section~\ref{se-light-bend}---even in the linear approximation.
In such cases the deflection angle $\alpha$ does not have a trivial dependence on the impact parameter $b$, which makes it not possible to define a meaningful generalized Eddington parameter $\gamma$, like in~\eqref{hatAlfa}, if the trajectory of the light ray comprises regions where $\chi_2$ does not have an approximate Newtonian form.
 
We have analysed both local (polynomial) and nonlocal (non--polynomial) models of gravity. The most interesting result was obtained in the case of analytic nonlocal gravity where we were able to perform a full analytic computation and cast the final expression of the deflection angle in terms of elementary functions. 
In this vein, it is  also useful to remark the efficiency of the method based on the Gauss--Bonnet theorem to evaluate the bending angle in the case of axisymmetric spacetimes~\cite{Gibbons:2008rj,Ono:2017pie}.

Before concluding let us emphasize that we only worked in the linearised regime, {\it i.e.,} weak--field and slow rotation, and neglected finite size effects. Therefore, future investigations are needed to extend our results to strong gravity regime with possible applications to the black hole shadow (see, for instance,~\cite{Jusufi:2019caq,Younsi:2016azx}). Indeed, this would be very interesting especially in light of the recent first ever captured image of a black hole by the Event Horizon Telescope Collaboration~\cite{Akiyama:2019cqa}.


\subsection*{Acknowledgements}
L.~B. acknowledges financial support from JSPS and KAKENHI Grant--in--Aid for JSPS Postdoctoral Fellows No.~JP19F19324.



\end{document}